\documentclass{article}
\usepackage{frascatiphys_net}

\usepackage{graphicx}
\usepackage{amsmath}
\usepackage{amsfonts}
\usepackage{amssymb}
\begin{document}
\title{
Kaonic Quantum Erasers at KLOE $2$: \\
{\em ``Erasing the Present, changing the Past''}}
 %Erasing
%the Past and impacting the Future with Neutral Kaons at KLOE $2$ }
\author{
Albert Bramon\\
{\em Grup de F{\'\i}sica Te\`orica, Universitat Aut\`onoma de
Barcelona,}\\ {\em E--08193 Bellaterra, Spain} \\
Gianni Garbarino      \\
{\em Dipartimento di Fisica Teorica, Universit\`a di Torino}\\{\em
and
INFN, Sezione di Torino, I--10125 Torino, Italy}\\
Beatrix C. Hiesmayr\\
{\em Institut f\"ur Theoretische Physik, Faculty of Physics,}\\\em{
University of Vienna, Boltzmanngasse 5, 1090 Vienna, Austria} }
\maketitle \baselineskip=11.6pt
\begin{abstract}
Neutral kaons are unique quantum systems to show some of the most
puzzling peculiarities of quantum mechanics. Here we focus on a
quantitative version of Bohr's complementary principle and on
quantum marking and eraser concepts. In detail we show that neutral
kaons $(1)$ are kind of double slit devices encapsulating Bohr's
complementarity principle in a simple and transparent way, and $(2)$
offer marking and eraser options which are \textbf{\em not} afforded
by other quantum systems and which can be performed at the
DA$\Phi$NE machine.
\end{abstract}
\baselineskip=14pt
%

%%%%%%%%%%%%%%%%%%%%%%%%%%%%%%%%%%%%%%%%%%%%%%%%%%%%%%%%%%%%%%%%%%%%%%%%%%%%%%%%%%%%%%%%%%%%%%%%
\section{Introduction}
%%%%%%%%%%%%%%%%%%%%%%%%%%%%%%%%%%%%%%%%%%%%%%%%%%%%%%%%%%%%%%%%%%%%%%%%%%%%%%%%%%%%%%%%%%%%%%%%

During the last fifteen years or so we have witnessed an interesting
revival of the research concerning some fundamental issues of
quantum mechanics. A very positive aspect of this revival is that it
has been driven by a series of impressive results which have been
possible thanks to improved experimental techniques and skillful
ideas of several experimental groups. As a result, some of the {\it
Gedankenexperimente} proposed and discussed in the earlier days of
quantum mechanics, or slight modifications of these proposals, have
been finally performed in the laboratory. Most of these experiments
belong to the fields of quantum optics and photonics; others make
use of (single) atomic or ion states. %Ã[cite soe work].

Among these kinds of experiments we concentrate on two types. The
first type concerns the old complementarity principle of Niels Bohr
for which a quantitative version became available in recent years.
%Ã\cite{GreenbergerYasin,Englert}.
This `quantitative complementarity' represents a major improvement
over older treatments and can be tested for rather simple quantum
systems. The second type of experiments requires more complex states
consisting of entangled two--particle systems. With these bipartite
states at hand one can test much more subtle quantum phenomena such
as the so called `quantum eraser', which adds puzzling space--time
considerations to the previous, Bohr's complementarity issue.

The aim of our contribution is to analyze the role that neutral
kaons can play in this two types of experiments. A copious source of
entangled neutral meson pairs, such as those produced in the Daphne
$e^+ e^-$ machine, can be shown to be extremely useful for this
purpose.

%%%%%%%%%%%%%%%%%%%%%%%%%%%%%%%%%%%%%%%%%%%%%%%%%%%%%%%%%%%%%%%%%%%%%%%%%%%%%%%%%%%%%%%%%%%%%%%%
\section{Kaons as double slits}\label{doubleslit}
%%%%%%%%%%%%%%%%%%%%%%%%%%%%%%%%%%%%%%%%%%%%%%%%%%%%%%%%%%%%%%%%%%%%%%%%%%%%%%%%%%%%%%%%%%%%%%%%

The famous statement  about quantum mechanics ``\textit{the double
slit contains the only mystery}'' of Richard Feynman is well known,
his statement about neutral kaons is not less to the point:
``\textit{If there is any place where we have a chance to test the
main principles of quantum mechanics in the purest way ---does the
superposition of amplitudes work or doesn't it?--- this is it}''
\cite{Feynman}. In this section we argue that single neutral kaons
can be considered as double slits as well.

Bohr's complementarity principle and the closely related concept of
duality in interferometric or double--slit like devices are at the
heart of quantum mechanics. The well--known qualitative statement
that ``\textit{the observation of an interference pattern and the
acquisition of which--way information are mutually exclusive}'' has
only recently been rephrased to a quantitative statement
\cite{GreenbergerYasin,Englert}:
\begin{eqnarray}\label{comp}
{\cal P}^2(y)+{\cal V}_0^2(y)\leq 1\;,
\end{eqnarray}
where the equality is valid for pure quantum states and the
inequality for mixed ones. ${\cal V}_0(y)$ is the fringe visibility,
which quantifies the sharpness or contrast of the interference
pattern (the ``wave--like'' property), whereas ${\cal P}(y)$ denotes
the path predictability, i.e., the \textit{a priori} knowledge one
can have on the path taken by the interfering system (the
``particle--like'' property). The path predictability is defined by
\cite{GreenbergerYasin}
\begin{equation}
{\cal P}(y)\;=\;|p_I(y)-p_{II}(y)|\;,
\end{equation}
where $p_I(y)$ and $p_{II}(y)$ are the probabilities for taking each
path ($p_I(y)+p_{II}(y)=1)$. Both ${\cal V}_0(y)$ and ${\cal P}(y)$
depend on the same parameter $y$ related somehow to the geometry of
the interferometric setup. It is often too idealized to assume that
the predictability and visibility are independent of this external
parameter $y$. For example, consider a usual experiment with a
vertical screen having a higher and a lower slit. Then the intensity
is generally given by
\begin{equation}
I(y)\;=\; F(y)\;\big(1+{\cal V}_0(y) \cos(\phi(y)\big)\;,
\end{equation}
where $F(y)$ is specific for each setup and $\phi(y)$ is the
phase--difference between the two paths. The variable $y$
characterizes in this case the position of the detector scanning a
vertical plane beyond the double--slit. An accurate description of
the interference pattern, whose contrast along a  wide scanned
region can hardly be constant, thus requires to consider the
$y$--dependence of visibility and predictability.

In Ref.~\cite{SBGH3} the authors investigated physical situations
for which the expressions of ${\cal V}_0(y), {\cal P}(y)$ and
$\phi(y)$ can be calculated analytically. This included interference
patterns of various types of double slit experiments ($y$ is linked
to position), as well as Mott scattering experiments of identical
particles or nuclei ($y$ is linked to a scattering angle). But it
also included particle--antiparticle oscillations in time due to
particle mixing, as in the neutral kaon system. In this case, $y$ is
a time variable indirectly linked to the position of the kaon
detector or the kaon decay vertex. Remarkably, all these two--state
systems, belonging to quite distinct fields of physics, can then be
treated via the generalized complementarity relation (\ref{comp}) in
a unified way. Even for specific thermodynamical systems, Bohr's
complementarity can manifest itself, see Ref.~\cite{HV}. Here we
investigate the neutral kaon case.

The time evolution of an initial $K^0$ state is given by
\begin{eqnarray}
\label{K-time-evolution} |K^0 (t)\rangle &=& \frac{1}{\sqrt{2}}
e^{-i m_L t-\frac{\Gamma_L}{2} t}\left\lbrace e^{i \Delta m
t+\frac{\Delta\Gamma}{2} t}\,  |K_S\rangle +
|K_L\rangle\right\rbrace\;,
\end{eqnarray}
where (here and in the following) inessential $CP$ violation effects
are safely neglected. In our notation $\Delta m \equiv m_L- m_S$ is
the (small) mass difference between the long-- and short--lived kaon
components whose time evolution is simply given by
\begin{eqnarray}
\label{KSL-time-evolution} |K_S (t)\rangle = e^{- im_S t}
e^{-{1\over 2} \Gamma_S  t}\,  |K_S\rangle \;, \:\; |K_L (t)\rangle
= e^{- im_L t} e^{-{1\over 2} \Gamma_L  t}\,  |K_L\rangle  \;.
\end{eqnarray}
Note that there are no oscillations in time between these two states
and that their decay rates are remarkably different, $\Gamma_S
\simeq  579 \Gamma_L$, so that we can write $\Delta\Gamma  \equiv
\Gamma_L-\Gamma_S \simeq -\Gamma_S \simeq -2.1 \Delta m<0$.

State (\ref{K-time-evolution}) can be interpreted as follows. The
two mass eigenstates $|K_S\rangle$ and $ |K_L\rangle$, i.e. the two
terms in the right hand side of eq.~(\ref{K-time-evolution}),
represent the two slits. At time $t=0$ both terms (slits) have the
same weight (width) and constructively interfere with a common
phase. As time evolves, the $K_S$ component decreases faster than
the other one and this can be interpreted as a relative shrinkage of
the $K_S$--slit making more likely the `passage' through the
$K_L$--slit. In addition, the norm of eq.~(\ref{K-time-evolution})
decreases with time as a consequence of both $K_S$ and $K_L$ decays,
an effect which could be mimicked by an hypothetical shrinkage of
both slit widths in real double--slit experiments. The analogy gets
more obvious if we eliminate this latter effect by restricting to
kaons which survive up to a certain time $t$  and are thus described
by renormalizing the state (\ref{K-time-evolution}):
\begin{eqnarray}
\label{singlekaon} |K^0 (t)\rangle &\cong&
\frac{1}{\sqrt{2\cosh(\frac{\Delta\Gamma}{2} t)}}\;
e^{-\frac{\Delta\Gamma}{4} t}\left\lbrace e^{i \Delta m
t+\frac{\Delta\Gamma}{2} t}\,|K_S\rangle +
|K_L\rangle\right\rbrace\;.
\end{eqnarray}

The probabilities for detecting on this state either a $K^0$ or a
$\bar K^0$ are given by
\begin{eqnarray}
P(K^0, t)&=& \left| \langle K^0 | K^0(t) \rangle \right|^2  =
\frac{1}{2} \biggl\lbrace
1+\frac{\cos(\Delta m t)}{\cosh(\frac{\Delta\Gamma}{2} t)}\biggr\rbrace\nonumber\\
P(\bar K^0, t)&=& \left| \langle \bar K^0 | K^0(t) \rangle \right|^2
= \frac{1}{2} \biggl\lbrace 1-\frac{\cos(\Delta m
t)}{\cosh(\frac{\Delta\Gamma}{2} t)}\biggr\rbrace\;,
\end{eqnarray}
showing the well--known strangeness oscillations. We observe that
the oscillating phase is given by $\phi(t)=\Delta m \,t$ and the
time dependent visibility by
\begin{eqnarray}\label{visibility}
{\cal V}_0(t)=\frac{1}{\cosh(\frac{\Delta\Gamma}{2} t)}\;.
\end{eqnarray}
The path predictability ${\cal P}(t)$, which in our kaonic case
corresponds to a ``which width'' information, can be directly
calculated from eq.~(\ref{singlekaon})
\begin{eqnarray}
\label{predictability} {\cal P}(t)=\left| P(K_S, t)-P(K_L,
t)\right|=\left|\;\frac{e^{\frac{\Delta \Gamma}{2} t} \;-\;
e^{-\frac{\Delta \Gamma}{2}t}}{2\cosh(\frac{\Delta\Gamma}{2} t)}\;
\right|=\left|\;\tanh\big(\frac{\Delta\Gamma}{2} t\big)\right|\;.
\end{eqnarray}

The expressions for predictability (\ref{predictability}) and
visibility (\ref{visibility}) satisfy the complementary relation
(\ref{comp}) for all times $t$
\begin{eqnarray}\label{kaon-complement-relation}
{\cal P}^2(t)+{\cal V}_0^2(t) \;=\;
\tanh^2\big(\frac{\Delta\Gamma}{2}t\big) +
\frac{1}{\cosh^2(\frac{\Delta\Gamma}{2}t)} \:=\; 1\;.
\end{eqnarray}
For time $t=0$ there is full interference, the visibility is
maximal, ${\cal V}_0(t=0)=1$, and we have no information about the
lifetime, ${\cal P}(t=0)=0$. This corresponds to the central part
(around the zero--order maximum) of the interference pattern in a
usual double slit scenario. For very large times, i.e. $t\gg
1/\Gamma_S$, the surviving kaon is most probably in a long lived
state $K_L$ and no interference is observed since we have almost
perfect information on  ``which width'' is actually propagating. For
times between these two extremes and due to the natural instability
of the kaons, we obtain partial ``which width'' information and the
expected interference contrast is thus smaller than one. However, the
full information on the system is contained in eq.~(\ref{comp}) and is
always maximal for pure states.
%The full description of a pure quantum system, such as our evolving $|K^0 (t)\rangle$ state, is
%contained in Eq.(\ref{comp}) taken with the equal sign.

The complementarity principle was proposed by Niels Bohr in an
attempt to express the most fundamental difference between classical
and quantum physics. According to this principle, and in sharp
contrast to classical physics, in quantum physics we cannot capture
all aspects of reality simultaneously and the available information
on complementary aspects is always limited.
%information content
Neutral kaons encapsulate indeed this peculiar feature in the very
same way as a particle having passed through a double slit.
%But kaons are double slits
But kaons are double slit devices automatically provided by Nature
for free!

%%%%%%%%%%%%%%%%%%%%%%%%%%%%%%%%%%%%%%%%%%%%%%%%%%%%%%%%%%%%%%%%%%%%%%%%%%%%%%%%%%%%%%%%%%%%%%%%
\section{Kaonic quantum eraser}\label{quantumeraser}
%%%%%%%%%%%%%%%%%%%%%%%%%%%%%%%%%%%%%%%%%%%%%%%%%%%%%%%%%%%%%%%%%%%%%%%%%%%%%%%%%%%%%%%%%%%%%%%%

Two hundred years ago Thomas Young taught us how to observe
interference phenomena with light beams.  Much more later,
interference effects of light have been observed at the single
photon level.  Nowadays also experiments with very massive
particles, like fullerene molecules, have impressively demonstrated
that fundamental feature of quantum mechanics \cite{Arndt}. It seems
that there is no physical reason why not even heavier particles
should interfere except for technical ones. In the previous section
we have shown that interference effects disappear if it is possible
to know the path through the double slit.
%the knowledge on the path through the double slit is the reason why interference is lost.
The `quantum eraser', a  gedanken experiment proposed by Scully and
Dr\"uhl in 1982 \cite{scully82}, surprised the physics community: if
that knowledge on the path of the particle is erased, interference
can be brought back again.

Since that work many different types of quantum erasers have been
analyzed and experiments have been performed with atom
interferometers \cite{Duerr} and with entangled photon pairs
\cite{Herzog,Kim,Tsegaye,Walborn,Trifonov,KimKim}. In most of them,
the quantum erasure is performed in the so-called ``delayed choice''
mode which  best captures the essence and the most subtle aspects of
the eraser phenomenon. In this case the meter, the quantum system
which carries the mark on the path taken, is a system spatially
separated from the interfering system which is generally called the
object system. The decision \textit{to erase or not} the mark of the
meter system ---and therefore the possibility \textit{to observe or
not} interference--- can be taken long after the measurement on the
object system has been completed. This was nicely phrased by
Aharonov and Zubairy in the title of their review article
\cite{AharonovZubairy} as ``erasing the past and impacting the
future''.

Here we want to present four conceptually different types of quantum
erasers for neutral kaons, Refs.~\cite{SBGH1,SBGH6}. Two of them are
analogous to erasure experiments already performed with entangled
photons, e.g. Refs.~\cite{Herzog,Kim}. For convenience of the reader
we added two figures sketching the setups of these experiments,
Fig.~\ref{photonActive} and Fig.~\ref{photonPassive}. In the first
experiment the erasure operation was carried out ``actively'', i.e.,
by exerting the free will of the experimenter, whereas in the latter
experiment the erasure operation was carried out ``partially
actively'', i.e., the mark of the meter system was erased or not by
a well known probabilistic law: photon reflection or transmission in
a beam splitter. However, different to photons the kaons can be
measured by an \textit{active} or a completely \textit{passive}
procedure. This offers new quantum erasure possibilities and proves
the
%very concept
underlying working principle of a quantum eraser, namely, sorting
events according to the acquired information.

For neutral kaons there exist two relevant alternative physical
bases. The first basis is the strangeness eigenstate basis $\{|
K^0\rangle , |\bar K^0 \rangle\}$. The strangeness $S$ of an
incoming neutral kaon at a given time $t$ can be measured by
inserting at the appropriate point of the kaon trajectory a thin
piece of high--density matter. Due to strangeness conservation of
the strong interactions between kaons and nucleons, the incoming
state is projected either onto $K^0$, by a reaction like $K^0
p\rightarrow K^+ n$, or onto $\bar K^0$, by other reactions such as
$\bar K^0 p\rightarrow \Lambda \pi^+$, $\bar K^0 n\rightarrow
\Lambda \pi^0$ or $\bar K^0 n\rightarrow K^- p$. Here the nucleonic
matter plays the same role as a two channel analyzer for polarized
photon beams. We refer to this kind of strangeness measurement,
which requires the insertion of that piece of matter, as an
\textit{active} measurement.

Alternatively, the strangeness content of neutral kaons can be
determined by observing their semileptonic decay modes. Indeed,
these semileptonic decays obey the well tested $\Delta S=\Delta Q$
rule which allows the modes
\begin{equation}
\label{semileptonic-decays} K^0(\bar s d) \to \pi^-(\bar u
d)\;+\;l^+\;+\;\nu_l \; , \; \; \bar{K}^0(s\bar d) \to\pi^+(u \bar
d)\;+\;l^-\;+\;\bar\nu_l ,
\end{equation}
where $l$ stands for $e$ or $\mu$, but forbids decays into the
respective charge conjugated modes. Obviously, the experimenter has
no control on the kaon decay, neither on the mode nor on the time.
The experimenter can only sort the set of all observed events in
proper decay modes and time intervals. We call this procedure,
opposite to the \textit{active} measurement procedure described
above, a \textit{passive}  strangeness  measurement.

The second basis $\{K_S,K_L\}$ consists of the short-- and
long--lived states having well defined masses $m_{S(L)}$ and decay
widths $\Gamma_{(S)L}$. We have seen that it is the appropriate
basis to discuss the kaon propagation in free space because these
states preserve their own identity in time,
eq.~(\ref{KSL-time-evolution}). Due to the huge difference in the decay
widths, the $K_S$'s decay much faster than the $K_L$'s. Thus in
order to observe if a propagating kaon is a $K_S$ or $K_L$ at an
instant time $t$, one has to detect at which time it subsequently
decays. Kaons which are observed to decay before $\simeq t + 4.8\,
\tau_S$ have to be identified as $K_S$'s, while those surviving
after this time are assumed to be $K_L$'s. Misidentifications reduce
only to a few parts in $10^{-3}$, see also Refs.~\cite{SBGH1,SBGH6}.
Note that the experimenter doesn't care about the specific decay
mode, he has to allow for free propagation and to record only the
time of each decay event. We call this procedure an \textit{active}
measurement of lifetime. Indeed, it is by actively placing or
removing an appropriate piece of matter that the strangeness (as
previously discussed) or the lifetime of a given kaon can be
measured. Since one measurement excludes the other, the experimenter has
to decide which one is actually performed and the kind of information
thus obtained.
%One measurement excludes the other, a decision has to be taken.

%Since the neutral kaon system violates the $CP$ symmetry (recall Section \ref{K-quantumstates})
%the mass eigenstates are not strictly orthogonal, $\langle K_S|K_L\rangle\neq 0$. However,
On the other hand, neglecting $CP$ violation effects ---recall that
they are of the order of $10^{-3}$, like the just mentioned $K_S ,
K_L$ misidentifications--- the $K_S$'s can be also identified by
their specific decay into  $2\pi$ final states ($CP=+$), while
$3\pi$ final states ($CP=-$)  have to be associated with $K_L$
decays. As before, we call this procedure a \textit{passive}
measurement of lifetime, since the kaon decay times and decay
channels (two vs three pions) used in the measurement are entirely
determined by the quantum nature of kaons and cannot be in any way
influenced by the experimenter.

\subsection*{$\qquad$\textbf{(a) Active eraser with \textit{active} measurements}}

\begin{figure}
\begin{center}
\includegraphics[width=200pt,
keepaspectratio=true]{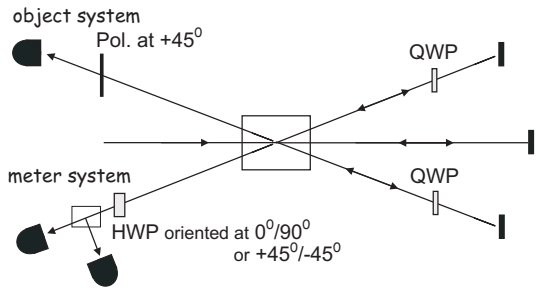} \caption{\it Sketched setup for an
active eraser. A bump beam transverses twice a non--linear crystal
producing photon pairs by (type II) parametric down--conversion. The
pairs produced in the first passage through the crystal (from left
to right) cross two times twice a quarter--wave plate (QWP) which
transforms an original horizontal polarized photon into a vertical
one and vice versa. The pairs produced in the second passage through
the crystal (from right to left) are superposed to the previous ones
and directed to the measurement devices. The signal or object photon
is always detected after crossing a polarization analyzer aligned at
$+45^\circ$. The idler or meter photon crosses a half--wave plate
(HWP) oriented at $0^\circ,90^\circ$ (first setup) or at
$\pm45^\circ$ (second setup) and is then analyzed by a polarization
beam splitter. In the first setup ---the meter photon is measured in
the $H/V$ basis--- one has full \textit{which way} information,
namely, one knows that the pair was produced at the first or second
passage. In the second setup ---the meter photon is measured in the
$+45^\circ/-45^\circ$ basis--- the information on the first or
second passage is erased. One then observes fringes for one--half of
the joint detections and the complementary (anti-)fringes for the
other half.} \label{photonActive}
\end{center}
\end{figure}

Let us first discuss the quantum eraser experiments performed with
photon pairs in Ref.~\cite{Herzog}. In this experiment (see
Fig.~\ref{photonActive}) two interfering two--photon amplitudes are
prepared by forcing a pump beam to cross twice the same nonlinear
crystal. Idler and signal photons from the first down conversion are
marked by subsequently rotating their polarization by $90^\circ$ and
then superposed to the idler (i) and signal (s) photons emerging
from the second passage of the beam through the same crystal. If
type--II spontaneous parametric down conversion were used, we would
had the two--photon state \footnote{The authors of
Ref.~\cite{Herzog} used type--I crystals in their experiment but this doesn't affect the present discussion.}
\begin{eqnarray}
\label{photonentangled} |\psi\rangle&=&
\frac{1}{\sqrt{2}}\biggl\lbrace \underbrace{|V\rangle_i
|H\rangle_s}_{\rm{second\;passage}} - \,e^{i \Delta\phi}\,
\underbrace{|H\rangle_i
|V\rangle_s}_{\rm{first\;passage}}\biggr\rbrace\;,
\end{eqnarray}
where $H$ and $V$ refer to horizontal and vertical polarizations
%and the symbol $\otimes$ for the tensor product of the states is dropped
%from now on.
The first and second terms in Eq.~(\ref{photonentangled}) correspond
to pair production at the second and first passage of the pump beam.
Their relative phase $\Delta \phi$, which depends on the difference
between the paths, is thus under control by the experimenter. The
signal photon, the object system, is always measured by means of a
two--channel polarization analyzer aligned at $+45^0$. Due to
entanglement, the vertical or horizontal idler polarization supplies
full \textit{which way} information for the signal system, i.e.,
whether it was produced at the first or second passage. In this
first experimental setup, where nothing is made to erase the
polarization marks, no interference can be observed in the
signal--idler joint detections. To erase this information, the idler
photon has to be detected also in the $+45^\circ/-45^\circ$ basis.
This is simply achieved in a second setup by changing the
orientation of the half--wave plate in the meter path.  Interference
fringes or, more precisely, fringes and anti--fringes can then be
observed in each one of the two channels when the relative phase
$\Delta \phi$ is modified.

In the case of entangled kaons produced by $\phi$ resonance decays
one starts with the state
\begin{equation}
\label{entangled} |\phi(0)\rangle  =  \frac{1}{\sqrt 2}\left[
|K^0\rangle_l |\bar{K}^0\rangle_r - |\bar{K}^0\rangle_l
|K^0\rangle_r\right] \simeq  \frac{1}{\sqrt 2}\left[ |K_L\rangle_l
|K_S\rangle_r - |K_S\rangle_l |K_L\rangle_r\right] ,
\end{equation}
where the $l$ and $r$ subscripts denote the ``left'' and ``right''
directions of motion of the two separating kaons and, as before,
CP--violating effects are neglected in the last equality. Kaons
evolve in time in such a way that the relevant state turns out to
depend on the two measurement times, $t_l$  and $t_r$, on the left
and the right hand side, respectively. More conveniently, this
two--kaon state can be made to depend only  on $\Delta t=t_l-t_r\,$
by  normalizing  to surviving kaon pairs\footnote{Thanks to this
normalization, we work with bipartite two--level quantum systems
like polarization entangled photons or entangled spin--$1/2$
particles. For an accurate description of the time evolution of
kaons and its implementation consult
Ref.~\cite{BertlmannHiesmayr2001}.}
\begin{eqnarray}
\label{timeentangled} |\phi(\Delta t)\rangle &=& \frac{1}{\sqrt
{1+e^{\Delta\Gamma \Delta t}}}\biggl\lbrace
|K_L\rangle_l|K_S\rangle_r - e^{i \Delta m \Delta t} e^{{1 \over 2}
\Delta
\Gamma \Delta t}|K_S\rangle_l|K_L\rangle_r\biggr\rbrace \\
&=& \frac{1}{2 \sqrt {1+e^{\Delta\Gamma \Delta t}}}
\left\{\big(1-e^{i \Delta m \Delta t} e^{{1 \over 2} \Delta \Gamma
\Delta t}\big)
\lbrace|K^0\rangle_l|K^0\rangle_r-|\bar K^0\rangle_l|\bar K^0\rangle_r\rbrace \right. \nonumber\\
&&\hphantom{\frac{1}{2 \sqrt {1+e^{\Delta\Gamma \Delta t}}}} \left.
+ \big(1+e^{i \Delta m \Delta t} e^{{1 \over 2} \Delta \Gamma \Delta
t}\big) \lbrace |K^0\rangle_l|\bar K^0\rangle_r-|\bar
K^0\rangle_l|K^0\rangle_r\rbrace \right\}\nonumber.
\end{eqnarray}
We note that the phase $\Delta m \Delta t$ introduces automatically
a time dependent relative phase between the two amplitudes.
Moreover, there is a complete analogy between the photonic state
(\ref{photonentangled}) and the two--kaon state written in the
lifetime basis, first eq.~(\ref{timeentangled}).

The marking and erasure operations can be performed on entangled
kaon pairs (\ref{timeentangled}) as in the optical case discussed
above. The object kaon flying to the left hand side is measured
always \textit{actively} in the strangeness basis, see
Fig.~\ref{QEkaon}(a). This active measurement is performed by
placing the strangeness detector at different points of the left
trajectory, thus searching  for oscillations along a certain $t_l$
range. As in the optical version, the kaon flying to the right hand
side, the meter kaon, is always measured \textit{actively} at a
fixed time $t_r^0$. But one chooses to make this measurement either
in the strangeness basis by placing a piece of matter in the beam or
in the lifetime basis by removing the piece of matter. Both
measurements are thus performed \textit{actively}. In the latter
case we obtain full information about the lifetime of the meter kaon
and, thanks to the entanglement, \textit{which width} the object
kaon has. Consequently, no interference in the meter--object joint
detections can be observed. This can be immediately seen from
eq.~(\ref{timeentangled}) once the left and right kaon kets are
written in the strangeness and lifetime bases, respectively. Indeed,
one obtains
\begin{eqnarray}
\label{probS}
&&P\left[K^0(t_l),K_S(t_r)\right]=P\left[\bar{K}^0(t_l),K_S(t_r)\right]
=\frac{1}{2\left(1+e^{\Delta\Gamma\Delta t}\right)} , \\
\label{probL}
&&P\left[K^0(t_l),K_L(t_r)\right]=P\left[\bar{K}^0(t_l),K_L(t_r)\right]
=\frac{1}{2\left(1+e^{-\Delta\Gamma\Delta t}\right)} ,
\end{eqnarray}
showing no oscillations in time. But interferences are recovered by
joint strangeness measurements on both kaons. From the last
eq.~(\ref{timeentangled}) one gets the following probabilities to
observe  like-- or unlike--strangeness events on both sides
\begin{eqnarray}
\label{lSprob}
P\left[K^0(t_l),K^0(t_r)\right]=P\left[\bar{K}^0(t_l),\bar{K}^0(t_r)\right]
= \frac{1}{4}\left[1-{\cal V}(\Delta t) \cos(\Delta m\, \Delta t)\right] , && \\
\label{uSprob}
P\left[K^0(t_l),\bar{K}^0(t_r)\right]=P\left[\bar{K}^0(t_l),K^0(t_r)\right]
= \frac{1}{4}\left[1+{\cal V}(\Delta t) \cos(\Delta m\, \Delta
\tau)\right] , &&
\end{eqnarray}
with a visibility
\begin{equation}
\label{visibility}
{\cal V}(\Delta t)=\frac{1}{\cosh(\Delta\Gamma\Delta t/2)}\,.
\end{equation}

\subsection*{$\qquad$\textbf{(b) Partially passive quantum eraser with \textit{active} measurements}}

\begin{figure}
\begin{center}
\includegraphics[width=270pt,keepaspectratio=true]{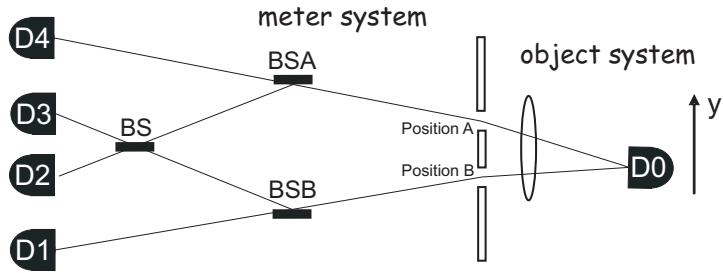} \caption{\it Sketched setup of a
`partially active' quantum  eraser. An entangled photon pair can be
produced either in region $A$ or in region $B$. If the detectors
$D1$ or $D4$ click, one knows the production region $A$ or $B$, i.e.
one has full \textit{which way} information also for its photon
partner. Clicks of the detectors $D2$ or $D3$ cannot contain this
information which has been erased at the central beam splitter.
Interference is observed only in this latter case. It is a
`partially active'  eraser, because the mark is erased by a
probabilistic law, however, the experimenter has still partially
control over the erasure, he can choose the ratio of transmittivity
to reflectivity of the beam splitter $BSA$ and $BSB$.}
\label{photonPassive}
\end{center}
\end{figure}

In Fig.~\ref{photonPassive} a setup is sketched where an entangled
photon pair is produced having a common origin in a region of points
including, e.g., points A and B. The experiment, realized in
Ref.~\cite{Kim} to which we refer for details, comprises a double
slit affecting the right moving object photon and a series of static
beam splitters and mirrors along the paths possibly followed by the
meter photon.  A look at Fig.~\ref{photonPassive} immediately shows
that ``clicks'' on detector $D1$ or $D4$ provide  ``which way''
information on this meter photon, which translates into the
corresponding information for its entangled, object partner. Joint
detection of these photon pairs shows therefore no interference. By
contrast, ``clicks'' on detector $D2$ or $D3$, which require the
cancelation of that ``which way''  information when the two possible
paths coincide on  the central beam splitter $BS$ in
Fig.~\ref{photonPassive}, lead to the expected, complementary
interference patterns for jointly  detected two--photon events
\cite{Kim}.

For neutral kaons, a piece of matter is permanently inserted into
both beams.  The one for the object photon has to be moved along the
left hand path in order to scan a certain $t_l$--range. The other
strangeness detector for the meter system is fixed on the right hand
path point corresponding to a fix $t_r^0$, see Fig.~\ref{QEkaon}(b).
The experimenter has to observe the region from the source to this
piece of matter at the right hand side. In this way the kaon moving
to the right ---the meter system--- takes the choice to show either
``which width'' information if it decays during its free propagation
before $t_r^0$ or not. In this latter case, it can be  absorbed at
time $t_r^0$ by the piece of matter. Therefore the lifetime or
strangeness of the meter kaon are measured \textit{actively}, i.e.,
distinguishing prompt and late decay events or $S= \pm 1$
kaon--nucleon interactions in matter. The choice whether the
``wave--like'' property or the ``particle--like'' property is
observed on the meter kaon is naturally given by the instability of
the kaons. It is ``partially active'', because the experimenter can
choose at which fixed time $t_r^0$ the piece of matter is inserted
thus making more or less likely the measurement of lifetime or
strangeness. This is analogous to the optical case where the
experimenter can choose the transmittivity of the two
beam--splitters $BSA$ and $BSB$ in Fig.~\ref{photonPassive}. Note
that it is not necessary to identify $K_S$ versus $K_L$ for
demonstrating the quantum marking  principle. The fact that this
information is somehow available is enough to prevent any
interference effects. These are recovered and oscillations reappear
if this lifetime mark is erased and joint events are properly
classified according to the measured strangeness of each kaon.

\subsection*{$\qquad$\textbf{(c) Passive eraser with ``\textit{passive}'' measurements on the meter}}

Next we consider the setup in Fig.~\ref{QEkaon}(c). We take
advantage ---and this is specific for kaons--- of the
\textit{passive} measurement. Again the strangeness content of the
object system
---the kaon moving to the left hand side--- is \textit{actively} measured by inserting a piece of matter into
the beam and thus scanning a given $t_l$ interval. In the beam of
the meter no material piece is inserted and  the kaon moving to the
right propagates freely in space. This corresponds to a
\textit{passive} measurement of either strangeness or lifetime on
the meter by recording the different decay \textit{modes} of neutral
kaons. If a semileptonic decay mode is found, the strangeness
content is measured and the lifetime mark is erased. The
distributions of the jointly  detected events will show the
characteristic interference fringes and antifringes. By contrast, if
a $\pi\pi$ or a $\pi\pi\pi$ decay is observed, the lifetime is
measured and thus ``which width'' information on the object system
is obtained and no interference is seen in the joint events. Clearly
we have a completely passive erasing operation on the meter, the
experimenter has no control whether the lifetime mark is going to be
read out or not.

This experiment is conceptually different from any other considered
two--level quantum system.

\subsection*{$\qquad$\textbf{(d) Passive eraser with ``\textit{passive}'' measurements}}

\begin{figure}
\begin{center}
\begin{large}{\textsf{\bf Kaonic quantum erasers}}\end{large}
\end{center}
\begin{flushleft}
(a) Active eraser with \textit{active} measurements \begin{tiny}(S: \textit{active}/\textit{active}; T: \textit{active})\end{tiny}\\
\begin{center}\includegraphics[width=180pt, keepaspectratio=true]{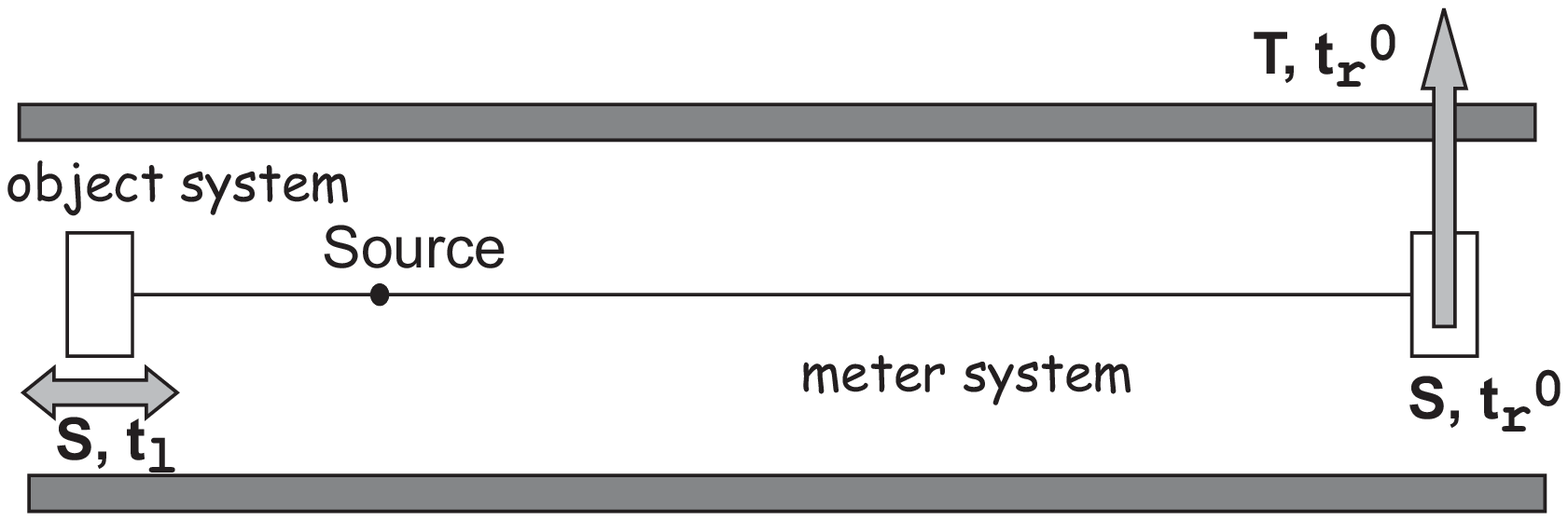}\end{center}
(b) Partially active eraser with \textit{active} measurements \begin{tiny}(S: \textit{active}/\textit{active}; T: \textit{active})\end{tiny}\\
 \begin{center}\includegraphics[width=180pt, keepaspectratio=true]{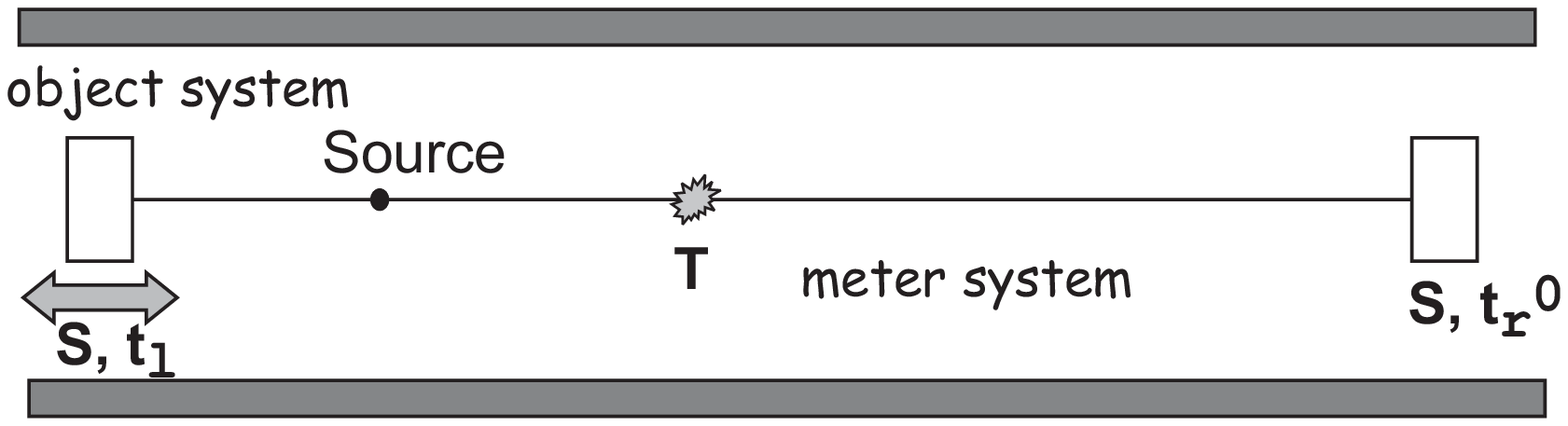}\end{center}
 (c) Passive eraser with \textit{passive} measurements
on the meter \begin{tiny}(S: \textit{active}/\textit{passive}; T:
\textit{passive})\end{tiny}\\
 \begin{center}\includegraphics[width=180pt, keepaspectratio=true]{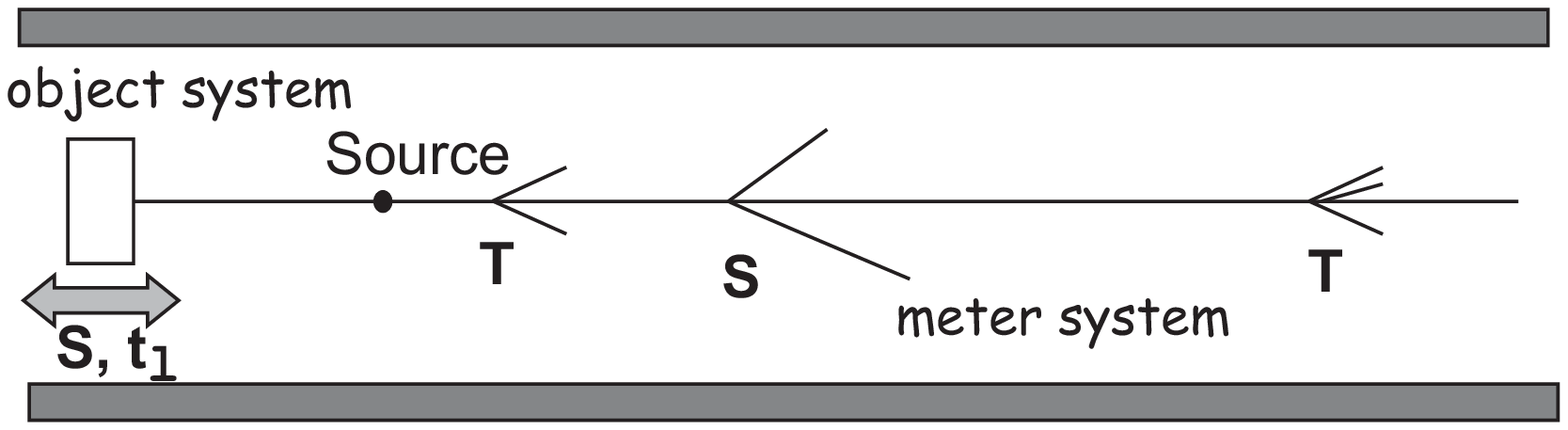}\end{center}
 (d) Passive eraser with \textit{passive} measurements
\begin{tiny}(S: \textit{passive}/\textit{passive}; T:
\textit{passive/passive})\end{tiny}\\
\begin{center}\includegraphics[width=180pt, keepaspectratio=true]{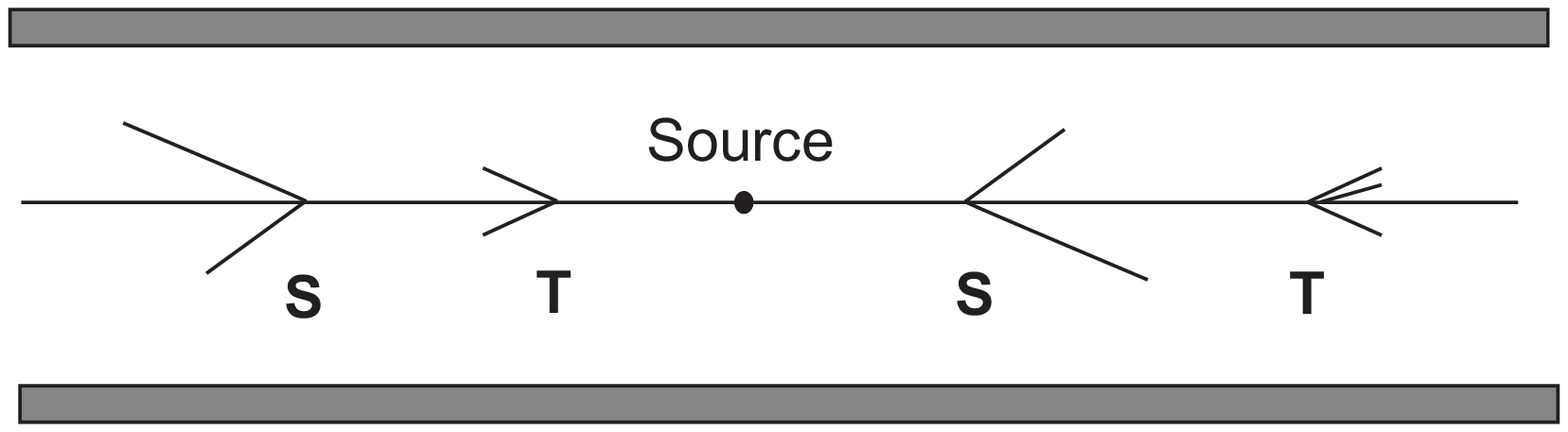}\end{center}
\caption{\em The figure shows four different setups for a quantum
marking and quantum erasing experiment. The first three, (a), (b)
and (c), have the object system on the left hand side on which the
strangeness is always \textit{actively} measured at time $t_l$. The
setups (a) and (b) are analogous to existing quantum eraser
experiments with entangled photons, see Fig.~\ref{photonActive} and
Fig.~\ref{photonPassive}. The setup (c) has no analog, because only
for kaons a \textit{passive} measurement is possible. For the last
setup, (d), it is not so clear which side plays the meter/object
role as it is totally symmetric and it involves only \emph{passive}
measurements.}\label{QEkaon}
\end{flushleft}
\end{figure}

Fig.~\ref{QEkaon}(d) sketches a setup where both kaons evolve freely
in space and the experimenter observes \textit{passively} their
decay modes and times. The experimenter has no control over
individual pairs neither on which of the two complementary
observables is measured on each kaon, nor when it is measured.

This setup is totally symmetric, thus it is not clear which side
plays the role of the meter. In this sense, one could claim that
this experiment should not be considered as a quantum eraser. But
one could also claim that this experiment reveals the true essence
of the erasure phenomenon: Until the two measurements (one in each
side) are completed, the factual situation is undefined; once one
has the measurement results on both sides, the whole set of joint
events can be classified in two subsets according to the kind of
information (on strangeness or on lifetime) that has been obtained.
The lifetime subset shows no interference, whereas fringes and
antifringes appear when sorting the strangeness subset events
according to the outcome, $K^0$ or $\bar K^0$, of the meter kaon.

Summarizing,
%it is remarkable that for all four presented setups
we have discussed four experimental setups combining \textit{active}
and \textit{passive} measurement procedures which lead to the same
observable probabilities. This is even true regardless of the
temporal ordering of the measurements, as follows immediately from
the fact that the $\Delta t$--dependent functions in
eq.(\ref{lSprob}), eq.(\ref{uSprob}) and in eq.(\ref{visibility}),
which govern the shape of the interference pattern, are even in this
variable $\Delta t = t_l -t_r$. Thus kaonic erasers can also be
operated in the ``\textit{delayed choice}'' mode as already
described in Ref.~\cite{SBGH6}. In our view this adds further light
to the very nature of the quantum eraser working principle: the way
in which joint detected events are classified according to the
available information. In the `delayed choice' mode, a series of
strangeness measurements is performed at different times $t_l$ on
the object kaons and the corresponding outcomes are recorded. Later
one can measure  either lifetime or strangeness on the corresponding
meter partner and, only now, full information allowing for a
definite sorting of each pair is available. If we choose to perform
strangeness measurements on the meter kaons and classify the joint
events according the $K^0$ or $\bar K^0$ outcomes, we complete the
information on each pair in such a way that oscillations and
complementary anti--oscillations appear in the corresponding
subsets. The alternative choice of lifetime measurements on meter
kaons, instead, does not give the suitable information to classify
the events in oscillatory subsets as before.
%%%%%%%
\section{Conclusions}
We have discussed the possibilities offered by neutral kaon states,
such as those copiously produced by $\phi$--resonance decays at
the DA$\Phi$NE machine, to investigate two fundamental issues of
quantum mechanics: quantitative Bohr's complementarity and quantum
eraser phenomena. In both cases, the use of neutral kaons allows for
a clear conceptual simplification and to obtain the relevant formulae
in a transparent and non--controversial way.

A key point is that neutral kaon propagation through the $K_S$ and
$K_L$ components automatically parallels most of the effects of
double slit devices.  Thanks to this,  Bohr's complementarity
principle can be quantitatively discussed in the most simple and
transparent way. Similarly, the relevant aspects of quantum marking
and the quantum eraser admit a more clear treatment with neutral
kaons than with other physical systems. This is particularly true
when the eraser is operated in the `delayed choice' mode and
contributes to clarify the eraser's working principle. Moreover, the
possibility of performing passive measurements, a specific feature
of neutral kaons not shared by other systems, has been shown to open
new options for the quantum eraser. In short, we have seen that,
once the appropriate neutral kaon states are provided as in the
DA$\Phi$NE machine, most of the additional requirements to
investigate fundamental aspects of quantum mechanics are
automatically offered by Nature for free.

The CPLEAR experiment \cite{CPLEAR} did only part of the job
(\textit{active} strangeness--strangeness measurements), but the
KLOE $2$ experiment could do the full program!

\vspace{1cm} \noindent {\bf Acknowledgement:} The authors thank the
projects SGR--994, FIS2005--1369 and EURIDICE HPRN-CT-2002-00311.
The latter allowed Beatrix Hiesmayr
 to work as a postdoc together with Albert Bramon and
Gianni Garbarino in Barcelona, where the idea of the kaonic quantum
eraser was born. We would also like to thank Antonio DiDomenico for
inviting us to the very interesting Frascati--Workshop ``Neutral
kaon interferometry at a $\Phi$--Factory: from Quantum Mechanics to
Quantum Gravity''.

%%%%%%%%%%%%%%%%%%%%%%%%%%%%%%%%%%%%%%%%%%%%%%%%%%%%%%%%%%%%%%%%%%%%%%%%%%%%%%%%%%%%%%%%%%%%%%%%
%%%%%%%%%%%%%%%%%%%%%%%%%%%%%%%%%%%%%%%%%%%%%%%%%%%%%%%%%%%%%%%%%%%%%%%%%%%%%%%%%%%%%%%%%%%%%%%%

%
\end{document}